\documentclass[onecolumn,preprintnumbers,amsmath,amssymb,preprint]{revtex4}
\usepackage{setspace}
\usepackage{dcolumn}
\usepackage{graphicx}
\usepackage{bm}
\usepackage[usenames,dvipsnames]{color} \usepackage{ulem}
\newcommand{\bk}{{\bf{k}}} 
\newcommand{\bq}{{\bf{q}}}

\newcommand{\pdag}{{\phantom\dagger}}

\newcommand{\ket}[1]{\left| #1 \right\rangle}

\begin{document}

\preprint{}
\title{Electron-lattice interactions strongly renormalize the charge transfer energy in 
the spin-chain cuprate Li$_\mathbf{2}$CuO$_\mathbf{2}$}

\author{Steve Johnston,$^1$ Claude Monney,$^{2,3}$ Valentina Bisogni,$^{4,5}$ Ke-Jin Zhou,$^{2}$  
Roberto Kraus,$^4$ G{\"u}nter Behr,$^4$ Vladimir N. Strocov,$^2$ Ji{\v r}i M{\'a}lek,$^6$ 
Stefan-Ludwig Drechsler,$^4$ Jochen Geck,$^4$ Thorsten Schmitt,$^2$ \& Jeroen van den Brink$^4$}
\vskip 1cm
\affiliation{%
$^1$Department of Physics and Astronomy, The University of Tennessee, Knoxville, TN 37996, USA\\
$^2$Paul Scherrer Institut, CH-5232 Villigen PSI, Switzerland\\
$^3$Institute of Physics, University of Zurich, Winterthurerstrasse 190, CH-8057 Zurich, Switzerland\\
$^4$Leibniz Institute for Solid State and Materials Research,  
  IFW Dresden, Helmholtzstrasse 20, D-01171 Dresden, Germany\\
$^5$National Synchrotron Light Source II, Brookhaven National Laboratory, Upton, NY 11973-5000, USA\\
$^6$Institute of Physics, ASCR, Na Slovance 2, CZ-18221 Praha 8, Czech Republic
}%

\date{\today}


\begin{abstract}
Strongly correlated insulators are broadly divided into two classes:
Mott-Hubbard insulators, where the insulating gap is driven by the Coulomb
repulsion $U$ on the transition-metal cation, and charge-transfer insulators,
where the gap is driven by the charge transfer energy $\Delta$ between the cation and
the ligand anions. The relative magnitudes of $U$ and $\Delta$ determine which class a
material belongs to, and subsequently the nature of its low-energy excitations.
These energy scales are typically understood through the local chemistry of the
active ions. Here we show that the situation is more complex in the
low-dimensional charge transfer insulator  Li$_\mathrm{2}$CuO$_\mathrm{2}$, where $\Delta$ has a large
non-electronic component. Combining resonant inelastic x-ray scattering with
detailed modeling, we determine how the elementary lattice, charge, spin, and
orbital excitations are entangled in this material. This results in a large
lattice-driven renormalization of $\Delta$, which significantly reshapes the
fundamental electronic properties of Li$_\mathrm{2}$CuO$_\mathrm{2}$.
\end{abstract}

\maketitle

\section{Introduction}   
The celebrated Zaanen-Sawatzky-Allen (ZSA) classification scheme \cite{ZSA} divides
strongly correlated insulators, such as transition metal oxides (TMOs), into
two broad categories: charge transfer or Mott-Hubbard insulators. Two
fundamental energy scales determine the boundary between these categories. The
first is the Coulomb repulsion $U$ associated with the transition-metal (TM)
cation site, which parameterizes the energy cost for ($d^\mathrm{n-1}d^\mathrm{n+1}$)-type charge
excitations. The second is the charge transfer energy $\Delta$ associated with
(d$^\mathrm{n-1}\underbar{L}$)-type charge excitations, where a hole moves from the cation site to the
ligand anions L. When these atomic energy scales dominate over electron
itinerancy, the emerging insulator is of the charge transfer type when $\Delta < U$ 
and of the Mott-Hubbard type when $\Delta > U$ \cite{ZSA}.

Determining which factors set the magnitude of these scales is important for
the most basic understanding of the behavior of TMOs. In an ionic picture, the
on-site Coulomb interaction $U$ sets the splitting of the lower and upper Hubbard
bands \cite{ZSA,OhtaPRB1991}, while the charge transfer energy 
$\Delta = \epsilon_\mathrm{p} - \epsilon_\mathrm{d}$ is typically set by the
relative electronegativity of the oxygen (O) anions and the ionization energy
of the TM cation \cite{OhtaPRB1991}. As such, copper oxides are typically classified as charge
transfer insulators, where their conduction band is derived from the copper (Cu) states
forming the upper Hubbard band, while the valence band is derived from the O $2p$ 
states. This dichotomy creates a fundamental asymmetry between electron and
hole doping processes, as reflected for example in the phase diagram of the
high-temperature superconducting cuprates \cite{ScalettarPRB1990,White}.

    Properly classifying a real material is a challenging task experimentally.
One needs to be able to determine the size of $\Delta$ and $U$ in the presence of
complications such as hybridization effects and 
additional interactions. Resonant inelastic x-ray scattering (RIXS) is a
powerful spectroscopic tool in this context \cite{AmentReview,KotaniReview}. 
It is capable of directly probing 
charge \cite{ChenPRL2010,OkadaPRB2001,MonneyPRL2013,Magnuson}, orbital \cite{SchlappaNature2012}, 
spin \cite{SchlappaPRL2009,BraicovichPRB2010,DeanNatureMaterials2013,LeTaconNaturePhysics2011}, 
and, as most recently
discovered, lattice excitations \cite{Yavas,AmentEPL2011,LeePRL2013,LeePRB2014}.
The observation of the latter is
particularly exciting, as RIXS can access the electron-phonon (e-ph) coupling
strengths directly \cite{AmentEPL2011}, and with element specificity \cite{LeePRL2013}.
This opens a direct means to study the influence of lattice dynamics on the 
fundamental electronic energy scales. 

In this work we perform such a study for the edge-shared charge transfer
insulator Li$_\mathrm{2}$CuO$_\mathrm{2}$ (LCO) in order to determine how the 
e-ph interaction helps to shape the charge transfer energy in this 
quasi-one-dimensional spin-chain cuprate. 
The active electronic degrees of freedom in LCO are formed from edge-shared
CuO$_\mathrm{4}$ plaquettes with a central Cu 3$d^\mathrm{9}$ cation \cite{Structure,KudoPRB2005,MizunoPRB1998}. 
As a result,
LCO harbors Zhang-Rice singlet (ZRS) charge excitons similar to those found in
the high-T$_\mathrm{c}$ cuprates \cite{OkadaPRB2001,MonneyPRL2013,LearmonthEPL2007}. 
The e-ph interaction is also expected to play a role
in this system. This was recently demonstrated for the related edge-shared
cuprate Ca$_\mathrm{2+x}$Y$_\mathrm{2-x}$Cu$_\mathrm{5}$O$_\mathrm{10}$ (CYCO), 
where charge carriers couple strongly to Cu-O
bond stretching phonon modes polarized perpendicular to the chain 
direction \cite{LeePRL2013,LeePRB2014}. We demonstrate here that a similar e-ph interaction occurs in
LCO. More importantly, however, we show that this interaction provides a
substantial contribution to $\Delta$, accounting for $\approx 54\%$ of its total
value. This result is obtained from a comprehensive analysis of high-resolution
oxygen K-edge RIXS \cite{KotaniReview,AmentReview} data that resolves individual phonon, $dd$, and charge
transfer excitations (including the ZRS exciton). This in turn allows us to
disentangle the elementary spin, charge, orbital, and lattice excitations over
an energy range of $\sim 10$ eV. If the e-ph interaction is omitted in our analysis,
the spectra imply a value $\Delta \approx 4.6$ eV; however, when the e-ph interaction is
properly accounted for, this value separates into a purely electronic
contribution of $\Delta_\mathrm{el} \approx 2.1$ eV and a very substantial phononic contribution of
about the same size $\Delta_\mathrm{ph} \approx 2.5$ eV. As such, the elementary excitations across
the charge transfer gap in LCO couple strongly to specific phonon modes,
enhancing their total energy cost. This result places the basic classification
of LCO in a new light, where the relevant energy scales are shaped not only by
the local chemistry of the atoms but also dynamically by interactions with
phonons that are relevant for many TMOs \cite{JohnstonPRB2010,MannellaNature2005,ShenPRL2004,MedardePRL1998}. 

\section{Results}
{\bf RIXS at the Oxygen K-edge in Li$_\mathbf{2}$CuO$_\mathbf{2}$} --- 
The oxygen K-edge RIXS process is sketched in Fig. 1. During the experiment,
photons with energy $\hbar\omega_\mathrm{in}$ and momentum 
$\hbar \bk_\mathrm{in}$ are absorbed by the system in its
initial state $\ket{i}$ via an O $1s \rightarrow 2p$ dipole transition. This creates an
intermediate state $\ket{m}$ with an O $1s$ core hole and an additional electron in
the conduction band. The resulting intermediate state then evolves in time
under the influence of the core-hole potential and the excited electronic
configuration. A number of elementary excitations are created in the system
during this time until the core hole decays, emitting an outgoing photon
(momentum $\hbar\bk_\mathrm{out}$ and energy $\hbar\omega_\mathrm{out}$) 
and leaving the system in an excited final state $\ket{f}$.

In order to understand how the e-ph interaction enters this process it is
important to examine further the states involved. The electronic ground state
in LCO, and other spin-chain cuprates, is largely of $\ket{i}_\mathrm{el} \sim 
\alpha\ket{d^\mathrm{9}}+\beta\ket{d^\mathrm{10}\underbar{L}}$ 
character, where $\underbar{L}$ denotes a hole on the ligand O. This state,
however, couples strongly to Cu-O bond-stretching phonons like the transverse
mode sketched in Fig. 1b. This coupling can occur in two ways. For instance, the
bond-stretching modes directly modulate the Cu-O hopping integral.
Alternatively, these modes can modulate the Madelung energy of the central Cu
atom, effectively lowering/raising the energy of the Cu site as the O atoms
move closer to/further from it. This latter mechanism cannot be effectively
screened in lower dimensions, and turns out to be the relevant coupling
mechanism for our analysis \cite{JvdBEPL2000,JohnstonPRB2010}. 
Since the electronic contribution to charge
transfer energy (in hole language) in this system is 
$\Delta_\mathrm{el} = \epsilon_\mathrm{p} - \epsilon_\mathrm{d}$, we can view
the phonon modes as modulating the charge transfer energy \cite{LeePRL2013}. This is confirmed
in Fig 1c, where we plot the linear variation in $\Delta_\mathrm{el}$ obtained from a static
point charge model under uniform expansions/compressions of the CuO plaquettes
in the direction perpendicular to the chain (see methods). 

The physical interpretation of this result is as follows. The lighter O atoms,
in an effort to eliminate the first order e-ph coupling and minimize the energy
of the system, shift to new equilibrium positions located closer towards the Cu
atoms. Subsequently, the new ground state of the system involves a coherent
state of phonon quanta $\left\{n_{\bq} \right\}$ that describes the distorted structure. The new
equilibrium positions also produce changes in the Madelung energy of the Cu
site, increasing the charge transfer energy in comparison to the value obtained
in the absence of the interaction. This renormalization of the charge transfer
energy is a bulk property of the crystal arising from the e-ph interaction with
the Cu 3$d^\mathrm{9}$ hole present in the ground state. As such, it will manifest in many
bulk spectroscopies including RIXS (this work), optical conductivity, and inelastic neutron
scattering (see Supplementary Note 1). It is important to note, however, that this
renormalization is inherently dynamic, as the oxygen atoms are free to respond
to changes in Cu hole density. This has observable consequences in the RIXS
spectra, as we now demonstrate. 

The RIXS process for LCO's initial state dressed by the phonon excitations is
sketched in Fig 1b. At low temperatures it is now predominantly $\ket{i} \sim \alpha
\ket{d^\mathrm{9},\left\{n_\bq\right\}} + \beta \ket{d^\mathrm{10}\underbar{L},\left\{n_\bq\right\}}$
in character. The intermediate state is formed
after the creation of a core hole on the O site, through an O $1s \rightarrow 2p$
transition. This creates an intermediate state of $\ket{m}\sim \beta\ket{d^\mathrm{10}p^6,
\left\{n_\bq \right\}}$ 
character, which corresponds to an upper Hubbard band excitation, where the
number of holes on the Cu site has changed. In response, the ligand O atoms
begin to relax towards new positions until the core hole decays. Ultimately,
this leaves the system in a final state with both excited electronic and
lattice configurations 
$\ket{f} \sim \alpha^\prime \ket{d^\mathrm{9},\left\{n^\prime_\bq\right\}} 
+ \beta^\prime \ket{d^\mathrm{10}\underbar{L},\left\{n^\prime_\bq\right\}}$. 

    It is important to stress that here the core hole provides us with a lens
through which we can view the e-ph interaction using RIXS. It does not generate
it. While the lattice excitations we probe are being generated in the
intermediate state, they carry information about the strength of the e-ph
interaction that is present in the initial and final states. The change in
carrier density introduced by the creation of the core hole excites the
lattice, but the way in which the lattice responds depends on strength and
details of the interaction. 

 \vspace{0.5cm}
{\bf Electron-phonon coupling in the RIXS data} --- 
The presence of the e-ph interaction in LCO is confirmed in our measured RIXS
spectra, shown in Fig. 2a. The x-ray absorption spectroscopy (XAS) spectrum (inset) has a prominent peak
centered at 529.7 eV, which corresponds to the discussed excitation into the
upper Hubbard band. The RIXS spectra, taken with incident photons detuned
slightly from this energy ($\hbar\omega_\mathrm{in} = 530.08$ eV, indicated by the arrow), are rich.
(Here we have shown data detuned from the UHB resonance, since the intensity of
the ZRS excitation is largest for this incident photon energy \cite{MonneyPRL2013}.) 
We observe a number of features, including a long tail of intensity extending from the
elastic line comprised of several phonon excitations; two nearly T-independent
peaks at $\sim 1.7$ and $\sim 2.1$ eV, which correspond to now well-known $dd$ 
excitations \cite{LearmonthEPL2007,HuangPRB2011};  
a T-dependent peak at $\sim 3.2$ eV, which corresponds to a
Zhang-Rice singlet excitation \cite{OkadaPRB2001,MonneyPRL2013}; 
and, finally, a band of charge transfer (CT)
excitations for $\hbar\Omega = \hbar\omega_\mathrm{out}-\hbar\omega_\mathrm{in}>4$. 
In this case, we are using the term CT excitation as
an umbrella term for any excitation where a Cu $3d$ hole has been transferred to
the O $2p$ orbitals, with the exception of the ZRS excitation. As such, CT
excitations include the florescence excitations. We have explicitly confirmed
each of these identifications by examining the character of the final state
wave functions obtained from our model calculations. 

The phonon excitations are more apparent in the high-resolution measurements of
the quasi-elastic and $dd$ excitation energy range, shown in Figs. 2d and 2f,
respectively. We observe clear harmonic phonon excitations separated in energy
by $\hbar\Omega_\mathrm{ph}\sim74$ meV, consistent with those reported for CYCO \cite{LeePRL2013,LeePRB2014}. 
This demonstrates that the e-ph coupling is a common phenomenon in the spin chain
cuprates. Another important aspect of the data is the positions of the ZRS and
CT excitations, which are determined by the charge transfer energy. From the
data we infer $\Delta \sim 4.6$ eV, which is significantly larger than the 3.2 eV
obtained from Madelung energy estimates based solely on local chemistry
considerations \cite{MizunoPRB1998}. This discrepancy can be accounted for by including the
bond-stretching phonons implied by the observed harmonic excitations in Figs 2d
and 2f.

 \vspace{0.5cm}
{\bf Electron-phonon contribution to the charge transfer energy} --- 
We assessed the phonon contribution to $\Delta$ by modeling the RIXS spectra within
the Kramers-Heisenberg formalism \cite{KotaniReview,AmentReview}. The initial, intermediate, and final
states were obtained from small cluster exact diagonalization (ED) calculations
that included the lattice degrees of freedom \cite{MonneyPRL2013,LeePRL2013}. The electronic 
model and its
parameters are the same as those used in a previous LCO study \cite{MonneyPRL2013}, however, we
have extended this model to include additional Cu $3d$ orbitals and kept the
bare charge transfer energy $\Delta_\mathrm{el} = \epsilon_\mathrm{p} - \epsilon_\mathrm{d}$ 
as a fitting parameter. This number
represents the size of the charge transfer energy in the absence of the e-ph
interaction. The model for the lattice degrees of freedom is similar to Ref. \cite{LeePRL2013}
but with an e-ph coupling strength parameterized by $g$ and the phonon energy
$\hbar\Omega_\mathrm{ph} = 74$ meV, as determined from our data (see methods). The calculated
spectra are shown in Fig. 2b, 2e, and 2g, where we have set $\Delta_\mathrm{el} = 2.14$ eV and
$g=0.2$ eV. This choice produces the best global agreement between the theory and
experiment both in terms of the positions of the CT and ZRS excitations, as
well as the intensities of the harmonic excitations in the $dd$ and quasi-elastic
regions. We also stress that the remaining parameters of the model were held
fixed during our fitting procedure, as their values are heavily constrained by
optical conductivity \cite{Drechsler,Malek}, EELS \cite{Drechsler}, 
inelastic neutron scattering \cite{LorenzEPL2009}, and RIXS (this
work and Ref. \cite{MonneyPRL2013}) measurements.

    A closer inspection of Fig. 2b reveals our main finding: the observed
positions of the ZRS and CT excitations are not set by the purely electronic
value of $\Delta_\mathrm{el}=2.14$ eV but rather the total $\Delta = \Delta_\mathrm{el}+\Delta_\mathrm{ph} \sim 4.6$ 
eV with $\Delta_\mathrm{ph}\propto \frac{g^2}{\hbar\Omega_\mathrm{ph}} \sim 2.5$ eV (see methods). 
In other words, the electron-lattice 
interaction is responsible for half of the effective charge transfer energy in
LCO. In order to stress this point, Fig. 2c shows results obtained from a
similar model where the e-ph interaction is taken out of the analysis. In order
to even qualitatively reproduce the positions of the ZRS and CT excitations in
this electronic-only model, $\Delta_\mathrm{el}$ must now be increased to 4.6 eV. Furthermore,
the phonon satellite peaks are absent in the quasi-elastic and $dd$ excitation
regions. This is a clear deficiency in the electronic-only model that is
corrected only when the e-ph interaction is included. It is clear that the
electron-lattice coupling plays a very significant role in establishing the
effective value of $\Delta$ in LCO.
    
We note that there is a small discrepancy between the theory and experiment;
namely, the relative intensity of the observed second phonon line with respect
to the first one is slightly stronger than the one captured by the cluster
calculation. While increasing the value of $g$ does increase the intensity of the
second phonon excitation relative to the first \cite{LeePRL2013}, the single mode model we have
adopted always produces a diminishing intensity in successive phonon
excitations. (We have also examined nonlinear e-ph interactions but these are
unable to account for this discrepancy.) We therefore speculate that increased
intensity in the second phonon excitation is due to multi-phonon processes that
cannot be included in our calculations due to the necessary truncation of the
phonon Hilbert space (see methods). For these reasons, we selected $g=0.2$ eV,
which is consistent with the Madelung energy considerations. This value also
provides a conservative estimate for the lattice contribution to $\Delta$.

    In Fig. 3 we compare the measured incident photon energy dependence to the
predictions of the e-ph coupled model as an additional verification. Here a
resonance behavior in the experimental data is observed, where the phonon
excitations emanating from the elastic line and $dd$ excitations persist to
higher energy losses as $\hbar\Omega_\mathrm{in}$ resonates with the upper Hubbard band excitation
in the XAS. Our experimental observations are in agreement with prior O $K$-edge
measurements on the related CYCO system \cite{LeePRL2013}. (In both materials, the observed
resonance behavior is damped with respect to similar behavior observed in gas
phase oxygen molecules \cite{HenniesPRL2005}. This is due to the increased number of core-hole
decay channels present in the solid \cite{LeePRL2013}.) Our model with the e-ph interaction
reproduces these features well. In contrast, the electronic model without the
electron-phonon coupling fails to capture these features. This underscores once
more the importance of the e-ph interaction for understanding the RIXS spectra
on even the qualitative level. 

\section{Discussion}
We have performed oxygen K-edge RIXS measurements on the edge-shared 1D cuprate
Li$_\mathrm{2}$CuO$_\mathrm{2}$, revealing clear phonon excitations in the RIXS spectra. These
excitations are well captured by a model that includes coupling to a Cu-O
bond-stretching optical phonon mode, which modulates the on-site energy of the
Cu orbitals and leads to a substantial renormalization of the effective charge
transfer energy. This renormalization is not a simple effect related to the
formation of the core hole. The non-zero e-ph interaction that we infer here is
present in the system regardless of the existence of the core hole. Thus the
corresponding renormalization of the charge transfer energy will also be
present in other spectroscopies such as optical conductivity \cite{MizunoPRB1998,Malek},  
EELS \cite{Drechsler}, and
inelastic neutron scattering \cite{LorenzEPL2009} (see Supplementary Note 1). 

    Our results show that the electron-phonon interaction is of relevance to
the Zaanen-Sawatzky-Allen classification of this material, where the lattice
contribution to the charge transfer energy accounts for nearly half of the
total value. Since the ensuing renormalizations can be very large in materials
possessing substantial electron-phonon couplings, we expect that such
considerations will prove to be important in other families of
quasi-one-dimensional correlated systems, where the lattice motion cannot be
effectively screened. For example, the related spin chain system CYCO likely
has a large lattice contribution to the charge transfer energy. 

\linespread{2}

\section{methods}
{\bf Sample preparation} --- 
Li$_2$CuO$_2$ samples were grown under elevated gas pressure (in a gas mixture
of Ar:O$_2$ with a ratio of 4:1 at the total pressure of 50 bar) in a vertical
travelling solvent floating zone facility with optical heating \cite{BehrLCO}. 
The powder for the feed rods of LCO was prepared by grinding and sintering LiOH
(Isotec, 99.9\% of $^7$LiOH powder was used) and CuO (Chempur 99.99\%) at 750
$^\circ$C. Because the powder was single phase after the first sintering, no
further annealing was done to avoid vaporization of lithium. The single phase
powder was pressed to polycrystalline rods (EPSI Engineered Pressure
Systems; 3500 bar) in latex tubes and sintered again at 800 $^\circ$C for 34 h.

 \vspace{0.5cm}
{\bf RIXS Measurements} --- The RIXS experiments 
were performed at the ADRESS beam line of the Swiss Light Source, Paul Scherrer
Institut, using the SAXES spectrometer \cite{Strocov,Ghiringhelli}. All spectra were recorded with
$\sigma$-polarized light in the scattering geometry shown in Fig. 1a (the scattering
angle was 130$^\circ$, with an incidence angle of 65$^\circ$). No momentum was transferred
into the system along the direction of the chain using this geometry. The
combined energy resolution was between 50 and 60 meV at the O $K$-edge
($\hbar\omega_\mathrm{in} \sim 530$ eV). About 150 photons were collected on the $dd$ 
excitations (maximum intensity) during 2 hours of data acquisition at an energy resolution
of 60 meV (RIXS spectra of Fig. 2a). About 300 photons were collected on the $dd$
excitations (maximum intensity) during 8 hours of data acquisition at an energy
resolution of 50 meV (RIXS spectra of Fig. 2 d and f). The samples were cleaved
in situ at a pressure of $\sim5\times10^{-10}$ mbar and a temperature $T = 20$ K. The surface
of the crystal was perpendicular to the $[101]$ axis such that the CuO$_\mathrm{4}$ 
plaquettes were tilted 21$^\circ$ from the surface.

 \vspace{0.5cm}
{\bf XAS and RIXS Intensities} --- 
The RIXS spectra at the O $K$-edge (1$s$ $\rightarrow$ 2$p$) 
were calculated using the Kramers-Heisenberg formula \cite{AmentReview,KotaniReview,KotaniPRB2002}.  
If the incoming and outgoing photons  
have energies (polarizations) $\hbar\omega_\mathrm{in}$ ($\hat{\mu}$) and  
$\hbar\omega_\mathrm{out}$ ($\hat{\nu}$), respectively, 
then the RIXS intensity is given by  
\begin{equation}\label{Eq:RIXS}
I_{\mu\nu}(\omega_\mathrm{in},\omega_\mathrm{out}) \propto 
\sum_f \left|\sum_{m} \frac{\langle f| D^\dagger_{\nu}|m\rangle \langle m| D^\pdag_{\mu}|i\rangle}{E_i +\hbar\omega_{in} - E_m + i\Gamma} 
\right|^2 \delta(E_i + \hbar\Omega - E_f). 
\end{equation}
Here, $\hbar\Omega = \hbar\omega_\mathrm{out}-\hbar\omega_\mathrm{in}$ is the energy loss;    
$|i\rangle$, $|m\rangle$, and $|f\rangle$ denote the initial, intermediate, 
and final states of the RIXS process, with eigenenergies $E_i$, $E_m$, and $E_f$, respectively; 
and $\Gamma$ is the lifetime of the core-hole, which we assume is independent of the intermediate state. 

The $1s \rightarrow 2p_\beta$ transition is induced by the 
dipole operator $D_\mu$. If no momentum is transferred to the sample ($\bq = 0$) by the 
incoming photon, then the dipole operator is given by 
\begin{equation}\label{Eq:Dipole} 
D_{\mu} = \sum_{i,\beta,\sigma} P_{\mu,\beta}[p^{\phantom\dagger}_{i,\beta,\sigma}s^\dagger_{i,\sigma} + h.c.],  
\end{equation} 
where $s^\dagger_{i,\sigma}$ ($s^{\phantom\dagger}_{i,\sigma}$)
creates (annihilates) a $1s$ core-hole of spin $\sigma$ on O site $i$; $p^\dagger_{i,\beta,\sigma}$ 
($p^{\phantom\dagger}_{i,\beta,\sigma}$) creates (annihilates) a spin $\sigma$ hole in the 
O 2$p_{\beta}$ orbital on the same site, and $P_{\mu,\beta} = \hat{\mu}\cdot \hat{r}_\beta$ 
is the projection of the photon polarization onto the orientation of the O 2p$_\beta$ orbital. 
For the scattering geometry shown in Fig. 1a, the transition operators are 
\begin{equation}
D_\sigma = \sum_{i,\sigma} [\cos(21^\circ)p^{\phantom\dagger}_{i,y,\sigma}s^\dagger_{i,\sigma} + h.c.]
\quad {\mathrm{and}} \quad
D_\pi = \sum_{i,\sigma} [\cos(65^\circ)p^{\phantom\dagger}_{i,x,\sigma}s^\dagger_{i,\sigma} + h.c.]
\end{equation}
for $\sigma$ and $\pi$-polarized light, respectively. (Note that the $p_z$ orbitals do not appear in 
these operators since we do not include them in our Hilbert space, see below.) Since the 
polarization of the outgoing photon was not measured in the experiment, the total intensity is 
given by an incoherent sum over outgoing polarizations 
$I_\sigma = \sum_\mu I_{\mu,\sigma}$ and $I_\pi = \sum_\mu I_{\mu,\pi}$.  

 \vspace{0.5cm}
{\bf Model Hamiltonian} --- 
The eigenstates $|i\rangle$, $|m\rangle$, and 
$|f\rangle$ were obtained from exact diagonalization (ED) of a 
small Cu$_3$O$_8$ cluster with an edge-shared geometry and open boundary 
conditions, as shown in Fig. 1b. The orbital basis contains the 
$3d_\mathrm{xy}$, $3d_\mathrm{x^2-y^2}$, and $3d_\mathrm{3z^2-r^2}$ 
orbitals on each Cu site, and the O $2p_\mathrm{x,y}$ orbitals on each O site. 
Throughout, $\alpha$ and $\alpha^\prime$ are used to index Cu orbitals, 
$\beta$ and $\beta^\prime$ are used to index O orbitals, and the Roman 
indices $i$, $j$ index the lattice sites. 

The full Hamiltonian is 
$H = H_\mathrm{o} + H_\mathrm{e-e} + H_\mathrm{ph} + H_\mathrm{e-ph}$,  
where $H_\mathrm{o}$ and $H_\mathrm{ph}$ contain the non-interacting 
terms for the electronic and lattice degrees of freedom, respectively, 
$H_\mathrm{e-e}$ contains the electron-electron interactions, and 
$H_\mathrm{e-ph}$ contains the electron-phonon (e-ph) interactions. 

The non-interacting terms for electronic degrees of freedom are  
\begin{equation}\label{Eq:Hkin}
H_\mathrm{o} = \sum_{i,\alpha,\sigma} \epsilon^d_{\alpha} d^\dagger_{i,\alpha,\sigma} d^\pdag_{i,\alpha,\sigma}
+ \sum_{i,\alpha,\sigma} \epsilon^p_{\beta} p^\dagger_{i,\beta,\sigma} p^\pdag_{i,\beta,\sigma}
+\sum_{i,j,\sigma,\alpha,\beta} t_{i,j}^{\alpha,\beta} d^\dagger_{i,\alpha,\sigma}
p^\pdag_{i,\beta,\sigma} + \sum_{i,j,\beta,\beta^\prime,\sigma} t_{i,j}^{\beta,\beta^\prime} 
p^\dagger_{i,\beta,\sigma} p^\pdag_{j,\beta^\prime,\sigma}, 
\end{equation}
where the Cu operators $d^\dagger_{i,\alpha,\sigma}$ ($d^\pdag_{i,\alpha,\sigma}$) and 
O operators $p^\dagger_{i,\beta,\sigma}$ ($p^\pdag_{i,\beta,\sigma}$) create 
(annihilate) a hole of spin $\sigma$ in orbital $\alpha$ (or $\beta$) on atomic site $i$. 
In Eq. (\ref{Eq:Hkin}) $\epsilon^d_\alpha$ and $\epsilon^p_\beta$  
are the on-site energies of the Cu and O orbitals, respectively, while  
$t_{i,j}^{\alpha,\beta}$ and $t_{i,j}^{\beta,\beta^\prime}$ are the Cu-O and O-O 
hopping integrals, respectively. 

The electron-electron interactions include the on-site inter- and intra-orbital on 
each Cu ($H_\mathrm{e-e}^\mathrm{d}$) and O ($H_\mathrm{e-e}^\mathrm{p}$) site, the nearest-neighbor Cu-O 
repulsion and exchange interactions ($H^\mathrm{pd}_\mathrm{e-e}$), and the nearest-neighbor Cu-Cu 
repulsion ($H^\mathrm{dd}_\mathrm{e-e}$). The Cu on-site interactions take the form 
\begin{eqnarray}\nonumber
H_\mathrm{e-e}^\mathrm{d}&=&\frac{U_\mathrm{d}}{2}\sum_{i,\alpha,\sigma\ne\sigma^\prime} 
d^\dagger_{i,\alpha,\sigma}d^{\phantom\dagger}_{i,\alpha,\sigma}
d^{\dagger}_{i,\alpha,\sigma^\prime}d^{\phantom\dagger}_{i,\alpha,\sigma^\prime} + 
\frac{U_\mathrm{d}^\prime}{2}\sum_{i,\sigma,\sigma^\prime,\alpha\ne\alpha^\prime} 
d^\dagger_{i,\alpha,\sigma}
d^{\phantom\dagger}_{i,\alpha,\sigma}
{d}^\dagger_{i,\alpha^\prime,\sigma^\prime} 
{d}^{\phantom\dagger}_{i,\alpha^\prime,\sigma^\prime} \\ 
&+&\frac{J_\mathrm{d}}{2}\sum_{i,\sigma,\sigma^\prime,\alpha\ne\alpha^\prime} d^\dagger_{i,\alpha,\sigma} d^\dagger_{i,\alpha^\prime,\sigma^\prime} 
d^{\phantom\dagger}_{i,\alpha,\sigma^\prime}d^{\phantom\dagger}_{i,\alpha^\prime,\sigma} 
+\frac{J_\mathrm{d}^\prime}{2}\sum_{i,\sigma\ne\sigma^\prime,\alpha\ne\alpha^\prime} d^\dagger_{i,\alpha,\sigma} d^\dagger_{i,\alpha,\sigma^\prime} 
d^{\phantom\dagger}_{i,\alpha^\prime,\sigma^\prime}d^{\phantom\dagger}_{i,\alpha^\prime,\sigma}. 
\end{eqnarray}
The form of on-site O interactions, $H^\mathrm{p}_\mathrm{e-e}$, is the same. 
The nearest-neighbor Cu-O interactions take a similar form 
$H_\mathrm{e-e}^\mathrm{pd} = \sum_{\langle i, j\rangle} H_{ij}^\mathrm{pd}$, 
where the sum is over nearest-neighbor Cu and O sites and 
\begin{eqnarray}\nonumber
H_{ij}^\mathrm{pd}&=& U_\mathrm{pd}\sum_{\sigma,\sigma^\prime,\alpha,\beta} 
d^\dagger_{i,\alpha,\sigma}d^{\phantom\dagger}_{i,\alpha,\sigma}
p^\dagger_{j,\beta,\sigma^\prime}p^{\phantom\dagger}_{j,\beta,\sigma^\prime}  
+K_\mathrm{pd}\sum_{\sigma,\sigma^\prime,\alpha,\beta} d^\dagger_{i,\alpha,\sigma} p^\dagger_{i,\beta,\sigma^\prime} 
d^{\phantom\dagger}_{i,\alpha,\sigma^\prime}p^{\phantom\dagger}_{i,\beta,\sigma} \\
&+&K_\mathrm{pd}^\prime \sum_{\sigma,\alpha,\beta} [d^\dagger_{i,\alpha,\sigma} d^\dagger_{i,\alpha,-\sigma} 
p^{\phantom\dagger}_{i,\beta,-\sigma}p^{\phantom\dagger}_{i,\beta,\sigma} + 
p^\dagger_{i,\beta,\sigma} p^\dagger_{i,\beta,-\sigma} 
d^{\phantom\dagger}_{i,\alpha,-\sigma}d^{\phantom\dagger}_{i,\alpha,\sigma} ].
\end{eqnarray}
Finally, the Cu-Cu nearest-neighbor repulsion is given by 
\begin{equation}
H_\mathrm{e-e}^\mathrm{dd} = U_\mathrm{dd}
\sum_{\langle i,j\rangle,\alpha,\alpha^\prime,\sigma,\sigma^\prime}  
d^\dagger_{i,\alpha,\sigma}d^\pdag_{i,\alpha,\sigma}
d^\dagger_{j,\alpha^\prime,\sigma^\prime}d^\pdag_{j,\alpha^\prime,\sigma^\prime}.
\end{equation}

For the lattice model $H_\mathrm{lat}$ and $H_\mathrm{e-ph}$, we considered a
single bond-stretching mode that compresses the Cu-O bond in the
direction perpendicular to the chain direction, as indicated by the arrows in
Fig. 1b. The reduction to a single phonon mode is required to maintain a
manageable Hilbert space for the problem; however, this approximation is
sufficient to describe the phonons in the related system 
Ca$_\mathrm{2+x}$Y$_\mathrm{2-x}$Cu$_\mathrm{5}$O$_\mathrm{10}$ \cite{LeePRL2013}. 
In principle, these bond-stretching phonons couple to the carriers in the chain
via two microscopic mechanisms: the first is via the direct modulation of the
interchain hopping integrals. The second is via a modification of the Cu site
energies. The magnitude of the former can be estimated from the distance
dependence of the atomic hopping parameters. The magnitude of the latter can be
estimated using an electrostatic point charge model for the Madelung
energies \cite{OhtaPRB1991,MizunoPRB1998}. 
We carried out such calculations using known structural data \cite{Structure} 
and obtained the distance dependence of $\epsilon_\mathrm{p}-\epsilon_\mathrm{d}$ 
(neglecting crystal field
effects) for static compressions of the CuO$_\mathrm{2}$ chain, as shown in Fig. 1b. The
results are shown in Fig. 1c, where we obtain an e-ph coupling strength $g \sim 0.24$ 
eV. Calculations were then carried out for both coupling mechanisms and the 
Cu site energy modulation was found to have the the largest impact on the
calculated RIXS spectra. We therefore neglected the modulation of the hopping
integrals here for simplicity and introduced a Holstein-like coupling to 
the Cu site energies. Within this model the Hamiltonian for the lattice degrees of freedom is  
\begin{equation}\label{Eq:e-ph}
H_\mathrm{ph} + H_\mathrm{e-ph} = \hbar\Omega_\mathrm{ph} b^\dagger b^\pdag + 
g\sum_{i,\alpha,\sigma} d^\dagger_{i,\alpha,\sigma}d^\pdag_{i,\alpha,\sigma}(b^\dagger + b^\pdag),
\end{equation}
where $b^\dagger$ ($b$) creates (annihilates) a phonon quanta of the compression mode. 
The hilbert space for the lattice degrees of freedom is truncated at a large number 
of allowed phonon quanta ($\sim 200 $). We have checked to ensure that our results are 
not significantly changed for further increases in this cut-off. 

Finally, when calculating the intermediate states in Eq. (\ref{Eq:RIXS}), the Hamiltonian 
is augmented with the appropriate terms describing the Coulomb interaction with the 
core-hole \cite{OkadaPRB2001}. Specifically, we add 
\begin{equation}\label{Eq:Corehole}
H_\mathrm{ch} = \sum_{i,\sigma} \epsilon^\pdag_\mathrm{1s} n^\mathrm{s}_{i,\sigma} + 
U_\mathrm{q} \sum_{i,\alpha,\sigma,\sigma^\prime} n^\mathrm{s}_{i,\sigma}n^\mathrm{p}_{i,\alpha,\sigma^\prime}, 
\end{equation}
where $n^\mathrm{s}_{i,\sigma} = s^\dagger_{i,\sigma}s^\pdag_{i,\sigma}$ is the number 
operator for the 1$s$ core level on O site $i$, $\epsilon_\mathrm{1s}$ is the 
energy of the O 1$s$ core-hole, and $U_\mathrm{q}$ is the core-hole potential. 

 \vspace{0.5cm}
{\bf Model parameters} --- 
The multi-band Hamiltonian has a number of parameters that can be
adjusted; however, we are constrained by multiple experimental probes. To this
end we have a well-established set given in Ref. \cite{MonneyPRL2013}, which simultaneously
reproduces high-energy features in the RIXS data \cite{MonneyPRL2013}, Cu-Cu exchange interactions
inferred from inelastic neutron scattering measurements \cite{LorenzEPL2009}, and optical
conductivity and EELS measurements \cite{Drechsler} in Li$_2$CuO$_2$. Given this level of
descriptive power, we adopt the same parameter set here. 
 
When the e-ph interaction is included in the calculation we take (in units of eV)  
$\epsilon^\mathrm{d}_\mathrm{xy} = 0$, $\epsilon_\mathrm{x^2-y^2}^\mathrm{d} = 1.7$, 
$\epsilon^\mathrm{d}_\mathrm{3z^2-r^2} = 2.1$, 
$\epsilon_\mathrm{x}^\mathrm{p} = 2.14$ and 
$\epsilon_\mathrm{y}^\mathrm{p} = 2.04$. 
The Cu-O hopping integrals are (in eV) 
$t_\mathrm{p_x,d_{xy}} = 0.740$, $t_\mathrm{p_y,d_{xy}} = 0.638$, 
$t_\mathrm{p_x,d_{x^2-y^2}} = 0.318$, $t_\mathrm{p_y,d_{x^2-y^2}} = 0.440$,
$t_\mathrm{p_x,d_{3z^2-r^2}} = 0.413$, and $t_\mathrm{p_y,d_{3z^2-r^2}} = 0.385$. 
The O-O hopping integrals are (in eV) 
$t_\mathrm{p_x,p_x} = 0.840$ (0.240) and $t_\mathrm{p_y,p_y} = 0.21$ (0.960), for 
hopping parallel (perpendicular) to the chain direction. 
The Hubbard and Hunds interactions 
for the Cu sites are given by the Racah parameterization \cite{Dagotto} with $A = 6.45$, 
$B = 0.25$, and $C = 0.35$. The O interactions are $U_\mathrm{p} = 4.1$, 
$J_\mathrm{p} = J^\prime_\mathrm{p} = 0.6$, and $U^\prime_\mathrm{p} = U - 2J_\mathrm{p}$. The extended interactions are 
$U_\mathrm{dd} = 0.4$, $U_\mathrm{pd} = 0.8$, $U^\prime_\mathrm{pd} = U_\mathrm{pd} - 2J_\mathrm{pd}$ 
and $J_\mathrm{pd} = 0.096$. The phonon energy is 
taken to be $\hbar\Omega_\mathrm{ph} = 74$ meV, and the e-ph coupling strength $g$ is taken as a 
variable. The core-hole parameters are $U_q = 4.3$ eV 
and $\Gamma = 150$ meV for the O $K$-edge.  

All of the parameters remain the same for the calculations performed without 
e-ph coupling with the exception of 
$\Gamma = 300$ meV, $\epsilon_\mathrm{x}^\mathrm{p} = 4.64$, 
and $\epsilon_\mathrm{y}^\mathrm{p} = 4.74$ eV. 
It should be noted that this parameter set assumes a larger value for
the charge transfer energy in comparison to Ref. \cite{MonneyPRL2013}, 
and fails to capture the phonon features in the RIXS data (Fig. 2c). To correct this, 
we take the bare charge transfer energy $\Delta_\mathrm{el} = 
\epsilon_\mathrm{p} - \epsilon_\mathrm{d}$ and
the bare e-ph interaction strength $g$ as fitting parameters and keep all other
model parameters to be the same as those listed above when the e-ph interaction is included. 
We therefore regard the charge transfer energy $\Delta$ used in Ref. 9 as an 
effective charge transfer energy, which includes the effects of the electron-phonon interaction. 

 \vspace{0.5cm}
{\bf Madelung Energies} --- 
The coupling to the phonon mode enters into our calculations to first order in 
displacement via the modulation of the Cu and O Madelung energies $V_\mathrm{Cu}$ and $V_\mathrm{O}$, 
respectively. The Madelung energy for a given site $i$ can be estimated using an ionic model, 
and is given by $V_i = \sum_{i\ne j} Z_j e^2/|\vec{r}_i-\vec{r}_j|$, where $Z_je$ is the 
formal charge associated with the atom at site $j$. Neglecting crystal field effects, 
the difference between the Cu and O site energies is related to the difference in Madelung energies 
$\Delta V_\mathrm{M} = V_\mathrm{O} - V_\mathrm{Cu}$ by \cite{OhtaPRB1991} 
\begin{equation}
\Delta = \epsilon_\mathrm{p} - \epsilon_\mathrm{d} = \frac{\Delta V_\mathrm{M}}{\epsilon} - I_\mathrm{Cu}(3) + 
A_\mathrm{O}(2) - \frac{e^2}{d}, 
\end{equation}
where $A_\mathrm{O}(2)$ is the second electron affinity of O, $I_\mathrm{Cu}(3)$ is the third ionization 
energy of Cu, $d$ is the Cu-O distance, and $\epsilon$ is the high-frequency 
dielectric constant. The distance dependence of $\Delta$ can be estimated by calculating 
$\Delta V_\mathrm{M}$ using the Ewald summation technique and the known structural data \cite{Structure}. 
Assuming $(I_\mathrm{Cu}(3) - A_\mathrm{O}(2) + \frac{e^2}{d}) = 10.88$ eV 
and $\epsilon = 3.3$, we arrive at $\Delta = 3.2$ eV for the 
experimental lattice parameters, in agreement with Ref. \cite{MizunoPRB1998}. This value, 
however, is substantially lower than the value inferred from our RIXS study if the electron-phonon 
interaction is excluded.  

In order to estimate the strength of the e-ph interaction, we  
performed calculations where the Cu-O plaquettes were compressed by a distance  
$u$ in the directions indicated by the arrows in Fig. 1b. The resulting 
distance dependence of $\Delta(u)$ is plotted in Fig. 1c, where 
a linear dependence of $\Delta$ occurs over a wide range of displacements. 
To capture this, we  
parameterize the Cu site energy as $\epsilon_\mathrm{d} + \frac{d\Delta}{du}u = \epsilon_\mathrm{d} +  
\sqrt{\frac{\hbar}{2M_\mathrm{O}\Omega_\mathrm{ph}}}\frac{\partial \Delta}{\partial u} 
(b^\dagger + b)$, where $M_\mathrm{O}$ is the mass of oxygen. 
This results in an electron-phonon coupling of the form given in Eq. (\ref{Eq:e-ph}) 
with $g = \sqrt{\frac{\hbar}{2M_\mathrm{O}\Omega_\mathrm{ph}}}\frac{\partial \Delta}{\partial u}$. 
A linear fit to $\Delta(u)$ (shown in Fig. 1c) gives $\frac{\partial\Delta}{\partial u} = 5.66$ eV/$\AA$, which yields 
$g \sim 0.24$ eV. It should be stressed that this value of $g$ 
is an estimate based on a point charge model, however, it gives us an idea of the 
expected coupling strength.  
 
\vspace{0.5cm}
{\bf Renormalization of the charge-transfer energy} ---
As discussed in the main text, in the ground state of the LCO chain the oxygen atoms 
will shift to new equilibrium positions in response to the linear electron-phonon coupling 
terms of the Hamiltonian. This situation can be qualitatively understood by 
introducing shifted phonon operators 
$a^\dagger =  b^\dagger + \langle n^\mathrm{d} \rangle \frac{g}{\hbar\Omega_\mathrm{ph}}$
and 
$a = b + \langle n^\mathrm{d} \rangle \frac{g}{\hbar\Omega_\mathrm{ph}}$, 
where $\langle n^\mathrm{d}\rangle = \sum_{i,\sigma} \langle n_{i,\sigma}^\mathrm{d} \rangle$ 
is the average number of holes on the Cu site in the ground state. 
These new operators yield a shifted atomic position given by 
$\langle u \rangle \propto \langle b^\dagger + b\rangle \propto -\frac{2g\langle n^\mathrm{d}\rangle}{\hbar\Omega\mathrm{ph}}$. 
This shift of position is responsible for the renormalization of the charge transfer energy.  
After this transformation is made the phonon and electron-phonon coupled terms of 
the Hamiltonian [Eq. (\ref{Eq:e-ph})] reduce to
\begin{equation}
H_\mathrm{ph} + H_\mathrm{e-ph} = \hbar\Omega_\mathrm{ph} a^\dagger a^\pdag  
+ g\sum_{i,\sigma} \left(n^\mathrm{d}_{i,\sigma} - \langle n^\mathrm{d}_{i,\sigma} \rangle\right) 
\left(a^\dagger + a^\pdag\right)
-\frac{2\langle n^d\rangle g^2}{\hbar\Omega_\mathrm{ph}} \sum_{i,\sigma}n^\mathrm{d}_{i,\sigma},  
\end{equation}
where we have dropped an overall constant.
The second term describes the coupling to the lattice in the new equilibrium position, which is 
proportional to the fluctuation in Cu charge density from its ground state value. The third term 
can be folded into the definition of the Cu site energy with 
$\epsilon_\mathrm{d}^\mathrm{eff} = \epsilon_\mathrm{d} - 
2\frac{\langle n^\mathrm{d}_{i,\sigma}\rangle g^2}{\hbar\Omega_\mathrm{ph}}$. This 
gives an effective charge transfer energy $\Delta_\mathrm{eff} = \Delta_\mathrm{el} + \Delta_\mathrm{ph}$ 
where $\Delta_\mathrm{el} = \epsilon_\mathrm{p} - \epsilon_\mathrm{d}$ and 
$\Delta_\mathrm{ph} = 2\frac{\langle n^\mathrm{d}_{i,\sigma}\rangle g^2}{\hbar\Omega_\mathrm{ph}}$.
From these considerations one can also see that no isotope effect is predicted 
for $\Delta_\mathrm{ph}$, since both $g^2$ and $\hbar\Omega_\mathrm{ph}$ are proportional to 
the inverse of the mass of oxygen.

 \vspace{0.5cm}
{\bf Acknowledgements} --- 
 The authors thank M. Berciu, T. P. Devereaux, W. S. Lee, B. Moritz, and
 G. Sawatzky for useful discussions. This research has been funded by the Swiss
 National Science Foundation and the German Science Foundation within the D-A-CH
 programme (SNSF Research Grant 200021L 141325 and Grant GE 1647/3-1). C.M. also
 acknowledges support by the Swiss National Science Foundation under grant
 number PZ00P2 154867. Further support has been provided by the Swiss National
 Science Foundation through the Sinergia network Mott Physics Beyond the
 Heisenberg Model (MPBH). J.G. gratefully acknowledge the financial support
 through the Emmy-Noether program of the German Research Foundation (Grant No.
 GE1647/2-1). The experiments were performed at the ADRESS beamline of the Swiss
 Light Source at the Paul Scherrer Institut. 

 \vspace{0.5cm}
 {\bf Author Contributions} ---  
 C.M., V.B., K.-J. Z., R. K., V. N. S., J. G. and T. S. performed the RIXS
 experiments, G. B. grew the single-crystalline samples. S.J., C.M., T.S.,
 J.M., S.-L.D. and J. v.d.B analysed the data and developed the model. S.J.
 performed the cluster calculations. J.v.d.B., J. G., T.S., S.J. and C.M. 
 conceived and managed the
 project. S.J. and J.v.d.B. formulated the manuscript with the assistance of all
 other coauthors.

 \vspace{0.5cm}
 {\bf Competing Interests} --- The authors declare that they have no
competing financial interests.

 \vspace{0.5cm}
 {\bf Correspondence} --- Correspondence and requests for materials should be addressed 
 to S. J. ~(email: sjohn145@utk.edu) or J. vdB ~(email:j.van.den.brink@ifw-dresden.de).

\newpage
\linespread{1}

\begin{figure}[t]
 \includegraphics[width=\textwidth]{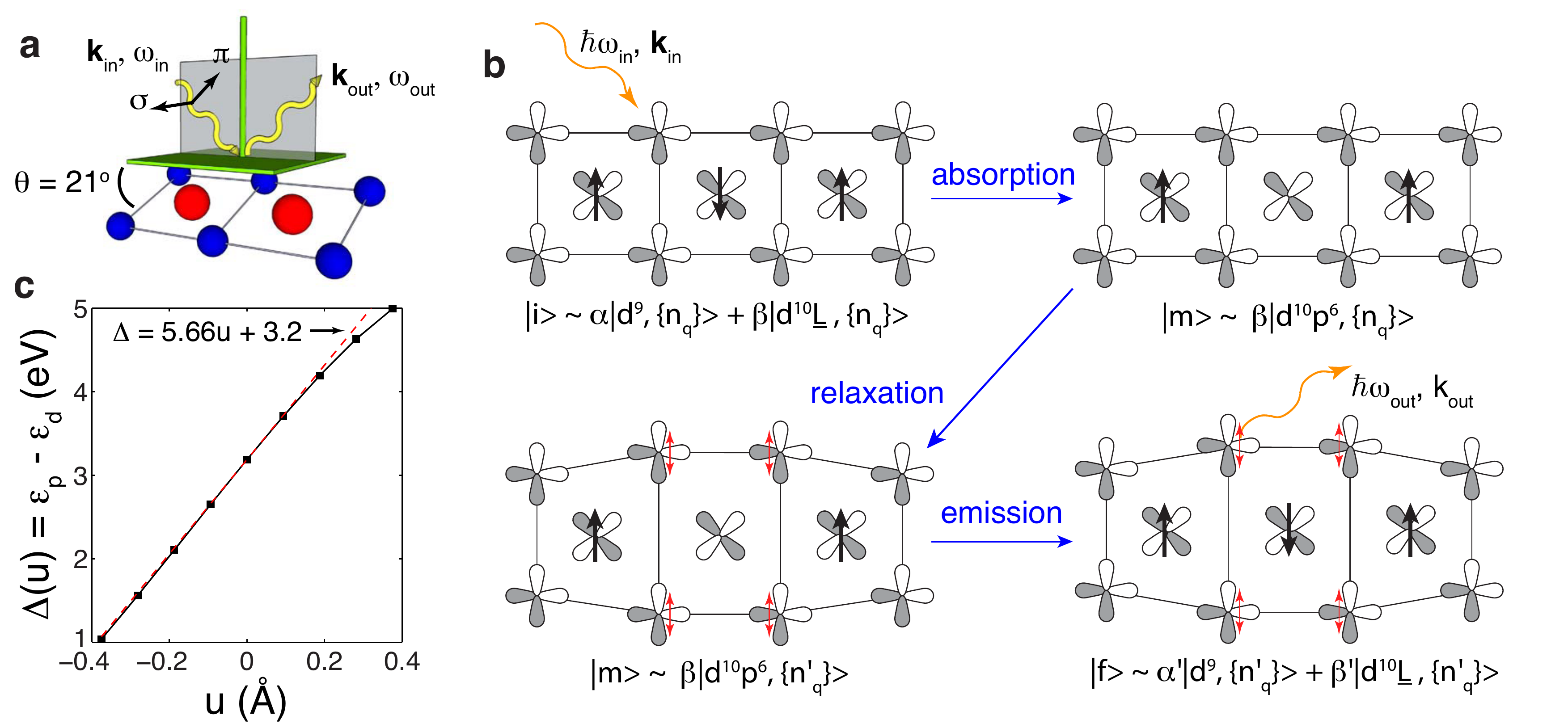}
 \caption{
 {\bf A cartoon sketch of the RIXS process}.
 {\bf a} A sketch of the experimental scattering geometry, showing the
 scattering plane (grey plane) perpendicular to the sample surface (green
 plane), making an angle of 21$^\circ$ with the CuO$_2$ chains, depicted here as
 a Cu$_2$O$_6$ dimer (Cu in red and O in blue).
 The wavy lines represent 
 the incoming and outgoing photons while the black arrows indicate the 
 polarization of the incoming photons with respect to the scattering plane. 
 {\bf b} 
 A sketch of the RIXS excitation process whereby the lattice is excited. The
 initial electronic state is predominantly of $\ket{i}_\mathrm{el}\sim \alpha \ket{d^9} 
 + \ket{d^{10}\underbar{L}}$ 
 character, where $\underbar{L}$ denotes a hole delocalized on the ligand oxygen 
 sites, while the initial lattice state involves a coherent state of 
 phonon quanta $\ket{\left\{ n_q\right\}}_\mathrm{ph}$ that describes the shifted equilibrium position 
 of the O atoms. The thick black arrows represent the low-temperature Ferromagnetic 
 spin structure of the Li$_\mathrm{2}$CuO$_\mathrm{2}$ chain.  
After the $1s \rightarrow 2p$ transition, an intermediate state of 
 $\ket{m}_\mathrm{el} \sim \beta  \ket{d^{10}}$
 character is formed, corresponding to an upper Hubbard band excitation
 where the number of holes on the Cu site has changed. Following this, the
 lattice relaxes in response to the change in Cu density, until the $1s$ core hole is filled,
 leaving the system in an excited electronic and lattice configuration 
 $\ket{\left\{ n_q^\prime\right \}}_\mathrm{ph}$. 
 The red arrows indicate the direction of the O atom's motion. 
{\bf c} The
 variation of the charge transfer energy $\Delta =\epsilon_\mathrm{p}-\epsilon_\mathrm{d}$ 
 as a function of a static
 compression $u$ of the Cu-O chains in a direction perpendicular to the chain
 direction. These values of $\Delta$ were obtained from an ionic picture and assuming a
 point charge model and should therefore be considered estimates. Crystal field
 effects have been neglected. The black points are the calculation 
 results while the red dashed line is a linear fit to the data at small displacement. 
 The linear electron-phonon coupling strength is
 related to the rate of change of $\Delta$ with respect to displacement with 
 $g=\sqrt{\frac{\hbar}{2M_\mathrm{O}
 \Omega_\mathrm{ph}}} \left( \frac{\partial\Delta}{\partial u}\right)$.
 }\label{Fig:1}
\end{figure}

\newpage
\begin{figure}
  \includegraphics[width=\textwidth]{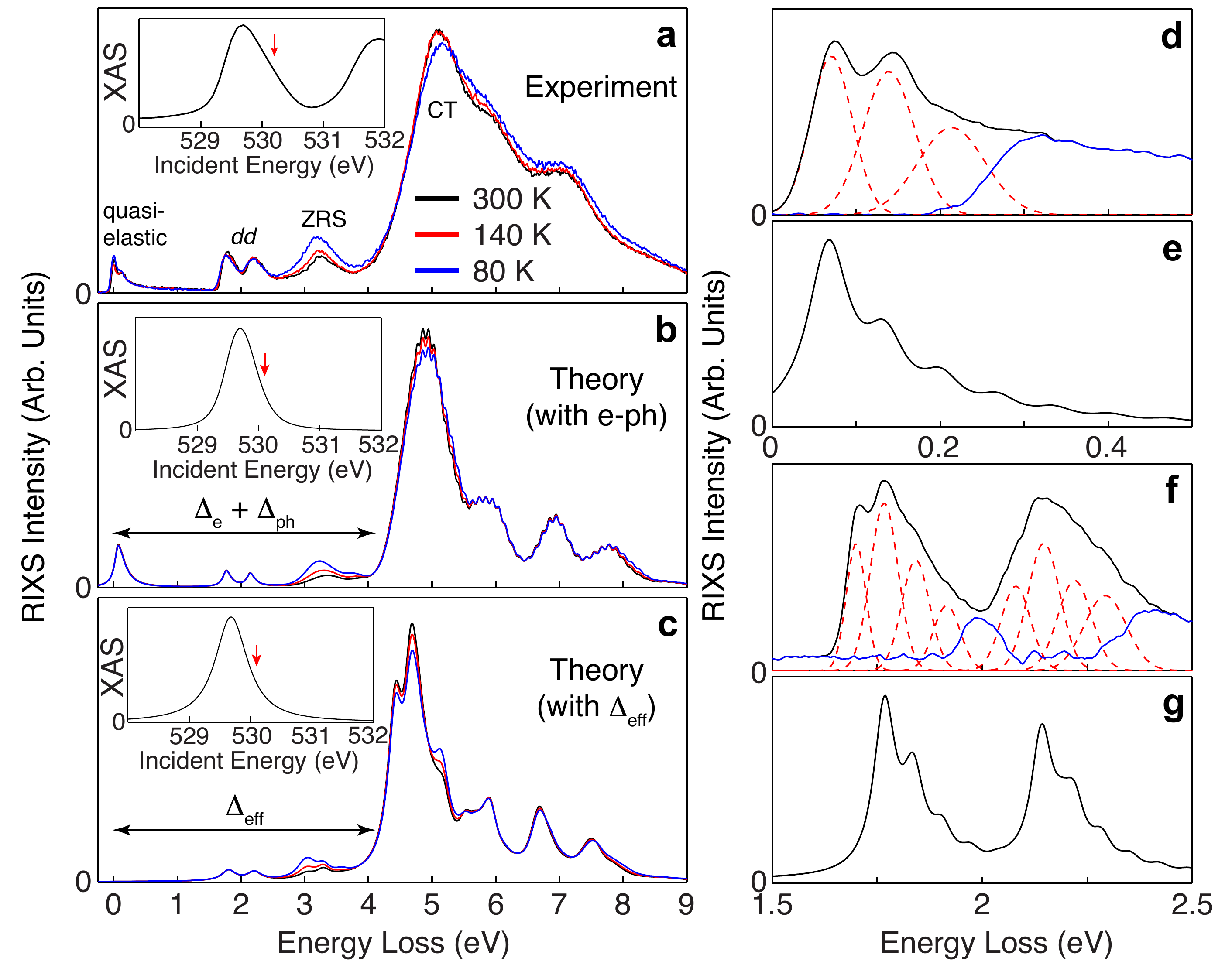}
 \caption{{\bf XAS and RIXS spectra of Li$_\mathrm{2}$CuO$_\mathrm{2}$ at the oxygen $K$-edge.}
 {\bf a} The measured RIXS spectra, recorded at various temperatures, as
 indicated. The incident photon energy for these measurements was detuned
 slightly from the upper Hubbard band peak in the XAS, as shown in the inset. 
 The incident phonon energy is indicated by the red arrow. 
 {\bf b}
 The calculated RIXS spectra obtained using a cluster model that includes
 coupling to the O-O bond-stretching mode. The calculated XAS spectrum is shown
 in the inset. For comparison, panel {\bf c} shows calculated spectra obtained from a
 model without coupling to the phonon mode and with an increased value of 
 $\Delta =  \epsilon_\mathrm{p}-\epsilon_\mathrm{d}\sim 4.6$ eV. 
 The detailed measured RIXS spectra highlighting the
 harmonic phonon excitations in the quasi-elastic and $dd$-excitation energy loss
 range are shown in panels {\bf d} and {\bf f}, respectively. 
 Here, the red dashed lines show Gaussian fits to the data that highlight the 
 individual phonon excitations. The blue line is the difference between the data 
 and the red dashed lines. The corresponding RIXS calculations
 are shown in panels {\bf e} and {\bf g}, respectively. In panels {\bf d} -- {\bf g} the incident photon
 energy coincides with the peak in the XAS intensity. Note that the elastic line
 has been removed from all of the calculated RIXS spectra for clarity. The
 spectra in panels {\bf e} and {\bf g} have been broadened using a Gaussian line shape with
 a standard deviation of 60 meV. In panels {\bf a} and {\bf b} this width was increased to
 130 meV in order to mimic additional broadening of CT features due to the bands
 formed by the O 2$p$ orbitals that are not well captured by our small Cu$_\mathrm{3}$O$_\mathrm{8}$ 
 cluster calculation.
 }\label{Fig:2}
\end{figure}

\newpage
\begin{figure}
  \includegraphics[width=0.5\textwidth]{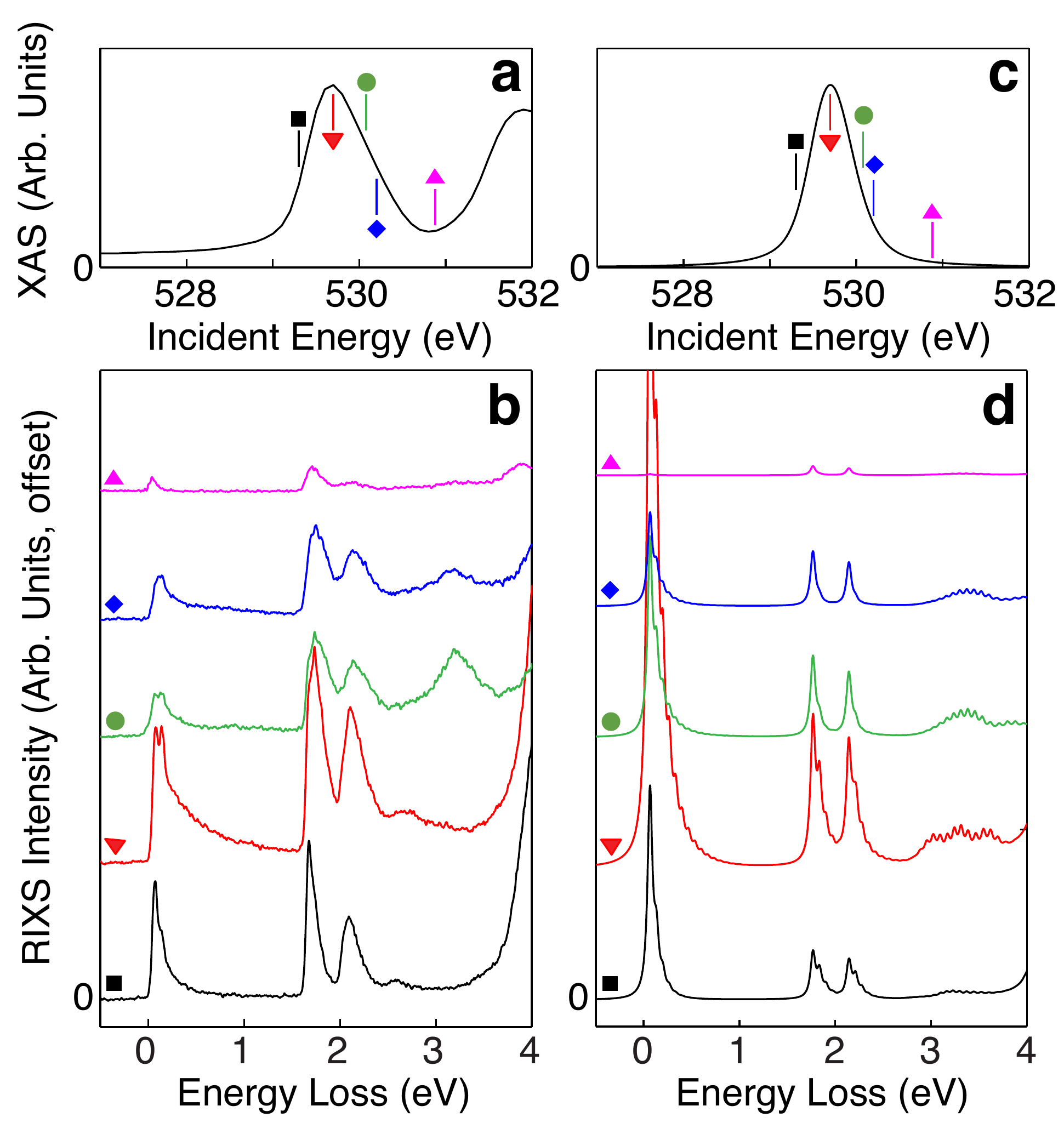}
  \caption{ {\bf The incident photon energy dependence of the RIXS spectra}. 
 Panels {\bf a} and {\bf c} show the measured and calculated XAS spectra,
 respectively. Calculations were performed using the model including coupling to
 the lattice. The measured and calculated RIXS spectra as a function of the
 incident photon energy are shown in panels {\bf b} and {\bf d}, respectively. 
 The RIXS spectra have been offset for clarity and the incident photon energy 
 is indicated by the color-coded symbols in the corresponding XAS plots. 
 The calculations have been broadened using a Gaussian line shape with a standard
 deviation of 60 meV.
\label{Fig:3}}
\end{figure}
 
\newpage

\end{document}